# Implementation of the Cluster Based Tunable Sleep Transistor Cell Power Gating Technique for a 4×4 Multiplier Circuit

Dipankar Saha[1], Subhramita Basak[1], Sagar Mukherjee[2], Sayan Chatterjee[1], C. K. Sarkar[1]

[1]Department of Electronics and Telecommunication Engineering, Jadavpur University,
Kolkata, West Bengal, India
[2]Department of Electronics and Communication Engineering, MCKV Institute of Engineering,
Liluah, Howrah, West Bengal, India
dipsah_etc@yahoo.co.in, basaksubhramita@gmail.com, sagarju87@gmail.com,
sayan1234@gmail.com, phyhod@yahoo.co.in

*Abstract*—**A modular, programmable, and high performance Power Gating strategy, called cluster based tunable sleep transistor cell Power Gating, has been introduced in the present paper with a few modifications. Furthermore, a detailed comparison of its performance with some of the other conventional Power Gating schemes; such as Cluster Based Sleep Transistor Design (CBSTD), Distributed Sleep Transistor Network (DSTN) etc.; has also been presented here. Considering the constraints of power consumption, performance, and the area overhead, while doing the actual implementation of any Power Gating scheme, it becomes important to deal with the various design issues like the proper sizing of the sleep transistors (STs), controlling the voltage drop (IR drop) across the STs, and obviously maintaining a desired performance with lower amount of delay degradation. With this notion, we tried to find out an efficient Power Gating strategy which can reduce the overall power consumption of any CMOS circuit by virtue of reducing the standby mode leakage current. Taking the different performance parameters into account, for an example circuit, which is actually the conventional 4×4 multiplier design, we found that the modified tunable sleep transistor cell Power Gating gives very much promising results. The reported architecture of the 4×4 multiplier with the tunable sleep transistor cell Power Gating, is designed using 45 nm technology and it consumes $1.3638 \times 10^{-5}$ Watt of Average Power while being operated with the nominal case of the bit configuration word, that is, "1000". At the same time, this design provides a delay of $2.5455 \times 10^{-10}$ second, which conveys a 2.29% improvement in the performance with respect to the best case delay as obtained in case of the conventional Power Gating scheme. The entire simulation work has been done using SPICE, whereas the results are obtained for a Supply Voltage ($V_{dd}$) of 1 Volt and a frequency of 200 MHz.**

*Keywords*- **multiplier; Power Gating; leakage power; sub-threshold current; delay; critical path; IR drop; delay degradation; CBSTD; DSTN; tunable sleep transistor cell**

## 1. Introduction

For the low leakage, high performance operation of any VLSI circuit, the Power Gating technique is treated as the most effective one which can substantially reduce the leakage current in standby mode. Now, considering the previously proposed circuit level approaches, the use of sleep transistors for Power Gating is found to be the most popular one [1-5]. When the circuit is in active mode these sleep transistors are 'ON'. But, for the standby mode of operation, these transistors get turned 'OFF', and that in turn disconnects the logic cells from the $V_{dd}$ (or, Ground) rail.

In conventional Power Gating architecture, a 'header' and a 'footer' switch used to be connected in series with the PUN (Pull-Up Network) and PDN (Pull-Down Network) of the logic circuits respectively. As illustrated in Fig. 1, the virtual-$V_{dd}$ rail (virtual-Ground rail) could be disconnected from the actual $V_{dd}$ (Ground) by turning-off the 'header' ('footer') sleep transistor; and thereby reducing the leakage power. But in active mode, these sleep transistors need to be turned 'ON', such that the logic circuit works fine as per its functionality. Now, instead of using both 'header' and 'footer' sleep transistors, the same leakage power reduction can be achieved by using any one of the two switches. Considering the perspective of area required, effective conductance etc., it is better to use NMOS sleep transistors as the footer switches [2, 3]. Now, for an effective implementation of Power Gating, to reduce leakage power, it is very much essential to determine the proper size of the sleep transistors. It is found, for a specific placement technique, the amount of performance degradation of the circuit usually depends on the size of the sleep transistors [1]. For the larger sleep transistors, it can be seen that the performance degradation is lesser [1]. But simultaneously those larger transistors require larger area, and a significant amount of driving energy [5]. Whereas, the insertion of smaller sleep transistors may cause an increase in performance degradation, which is also not acceptable [1]. So obviously, there is a trade-off in between the power consumption and the performance of the circuit.

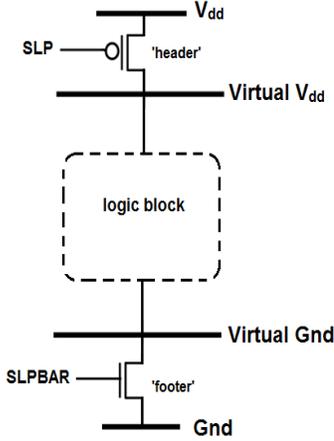

Figure 1. Conventional Power Gating architecture

To find out an effective Power Gating Strategy, a rigorous analysis has been done here, in this work. We started with the conventional Power Gating, where there used to be a large single transistor which can gate the entire logic circuit [6, 7]. Then we have considered another popular and well practiced Power Gating technique, called Cluster Based Sleep Transistor Design; and tried to find out its effectiveness in reducing the leakage power, as well as maintaining the performance of a logic circuit. After that, the concept of Distributed Sleep Transistor Network has been employed for the very same purpose. And lastly, a modified architecture of the tunable sleep transistor cell has been introduced to reduce the standby leakage of a logic circuit, without degrading the overall performance much. Now, for all the cases, the different Power Gating Strategies (as mentioned above) have been implemented on a basic circuit which is actually the 4×4 multiplier design as it is described in [8].

## 2. Leakage Power & McCMOS Technique

Though the reduction of the device dimensions, with each technology node, has increased the integration density as well as resulted in a substantial improvement of the speed [9, 10]; but unfortunately, considering the aspect of power consumption, this has led to a situation where the leakage power has become a major contributor to the total power consumption. Considering the deep-submicron devices or, the nano-devices, where the $v_{th}$ is quite low, the leakage power dissipation that occurs in a circuit, is mainly due the sub-threshold and the gate leakage current. Besides, the Gate Induced Drain Leakage (GIDL), the Band To Band Tunneling (BTBT) etc. are the other contributors which have become a concern in case of the advanced MOS devices [10].

Due to the non-zero minority carrier concentration, in the 'weak-inversion' region, there occurs a current conduction between the source and the drain of the MOS device; even if the applied gate voltage is below the $v_{th}$. This is actually the sub-threshold current [9]. Considering the 'weak-inversion', the DIBL (Drain Induced Barrier Lowering) effect as well as the body effect, we can model the sub-threshold current conduction as [9, 11],

$$I_{sub} = A \times e^{\frac{1}{mv_T}(V_G - V_S - v_{th0} - \gamma' V_S + \eta V_{DS})} \times \left(1 - e^{-V_{DS}/v_T}\right).$$

(1)

Where,

$$A = \mu_0 C_{ox}(W/L_{eff}) \times (v_T)^2 \times e^{1.8} \times e^{-\Delta v_{th}/\eta v_T},$$

$\mu_0$ is the zero bias mobility, $C_{ox}$ is the per unit area gate oxide capacitance, $v_{th0}$ is the zero bias threshold voltage, and $v_T$ is the thermal voltage. The sub-threshold swing co-efficient is denoted by $m$, whereas the linearized body effect co-efficient and the DIBL co-efficient are represented by the terms $\gamma'$ and $\eta$ respectively. And as there exists an exponential relationship of the sub-threshold current to the change in $v_{th}$, therefore assigning the higher-$v_{th}$ to the transistors in a circuit can be very useful in reducing the leakage current, and thereby reducing the leakage power [12]. But, the problem is that the higher-$v_{th}$ increases the equivalent ON-resistance ($R_{ON}$) for the transistors, and that in turn increases the delay [12]. The propagation delay through a transistor is generally denoted as,

$$T_{delay} = \frac{C_L V_{dd}}{K \cdot (V_{dd} - V_{th})^\alpha}.$$ (2)

Where, $K$ is a factor which depends on the gate size, as well as on the process. $\alpha$ takes any value between 1 and 2 depending on channel length [13]. Therefore, we can see that the reduction of the $v_{th}$ can be useful to improve the overall performance at low supply voltages [14]. But, as we reduce the $v_{th}$ of the transistor, leakage current starts playing a dominant role [11, 14]. Thus, maintaining the performance of the circuit as well as reducing the leakage power dissipation becomes a key challenge for designing any low-voltage, low power digital circuit.

For the conventional CMOS technology, the multiple channel length CMOS (McCMOS) technique is known to be one of the popular means by which we can reduce the leakage power [15]. As per the technique, the channel length of the transistors used in a circuit can be increased, wherever it is needed to control the leakage current. On the other hand, wherever it is required to maintain the performance (specially, for the transistors in critical path), we need to increase the width of the transistors [15]. Though, the use of McCMOS leads to an increase in the area overhead as well as the switched capacitance; but those are not that vulnerable. In comparison to the other leakage control techniques such as, multi-threshold-voltage, dual-threshold-voltage technique, body biasing technique etc., the McCMOS technique neither requires the additional processing steps nor, any additional biasing circuitry, thereby provides a quite simple but effective way for reducing the leakage in any CMOS circuit [15].

## 3. Power Gating Strategies

### 3.1 Conventional Power Gating

In case of conventional Power Gating, generally, there used to be a large single transistor (of width W) which can gate the entire logic circuit [6, 7]. In active mode, when the sleep transistor is 'ON', it provides a resistance $R_{ON}$, which is basically the channel resistance of the transistor. Let, $d$ be the 50% propagation delay for any logic block residing in a typical row of a CMOS circuit; whereas the load capacitance for the logic block and the supply voltage for the entire CMOS circuit are denoted by $C_L$ and $V_{dd}$ respectively. Then we can write,

$$d \propto \frac{C_L V_{dd}}{(V_{dd} - v_{th})^{\alpha}}, \quad (3)$$

where, $\alpha$ is the velocity saturation index [1]. Again, after the insertion of the ST, say the propagation delay value changes to $d'$. Thus,

$$d' \propto \frac{C_L V_{dd}}{(V_{dd} - v_{ST} - v_{th})^{\alpha}}, \quad (4)$$

where, $v_{ST}$ denotes voltage drop across the ST [1]. Now, the increase in delay $\Delta d$ can be denoted as,

$$\Delta d = (d' - d) \propto \frac{v_{ST}}{V_{dd} - v_{th}} d. \quad (5)$$

Therefore, the $\Delta d / d$ ratio (which is the ratio denoting the delay degradation of the logic block) is actually proportional to the $v_{ST}$ [1].

As per one of the mostly practiced methods, a constraint guaranteeing that $v_{ST}$ should not exceed 10 % of $V_{dd}$, must be met while sizing the ST for the logic block. If we represent $v_{ST}$ as a fraction of the supply voltage, while $I_{ST}$ is the maximum value of discharge current that flows through the sleep transistor during the active mode of operation, then the channel resistance of the sleep transistor can be represented as [5],

$$R_{ST} = \frac{V_{dd} \alpha_{drop}}{I_{ST}}. \quad (6)$$

Where, $V_{dd} \alpha_{drop}$ (which is, actually the fraction of supply voltage) denotes the maximum allowed voltage drop across the sleep transistor. Considering that the sleep transistor is operating in resistive region, we can further estimate the size of the transistor as [4, 5],

$$\left(\frac{W}{L}\right)_{ST} = \frac{\beta}{R_{ST}}. \quad (7)$$

Where, $\left(\frac{W}{L}\right)_{ST}$ is denoting the aspect ratio of the sleep transistor (L is the gate length, W is the width). And $\mu_0$ being carrier mobility, $C_{ox}$ being the oxide capacitance, $\beta$ can be expressed as [4],

$$\beta = \frac{1}{\mu_0 C_{ox}(V_{dd} - v_{th} - V_{dd} \alpha_{drop})}. \quad (8)$$

Therefore, we can see that the different physical parameters, like carrier mobility, threshold voltage, gate length etc. play an influential role in determining the size of the sleep transistor [5].

### 3.2 Cluster Based Sleep Transistor Design

A large sleep transistor with a greater value of W generally causes a significant area overhead; and that in turn results in an excess consumption of power. Furthermore, a larger sleep transistor may nullify the leakage power savings as the sleep transistor itself will contribute a considerable amount of leakage in standby mode [4]. To mitigate the aforesaid problem, we may go for a Cluster Based Sleep Transistor Design, where the different logic gates inside a circuit module, can be grouped into more than one clusters; and the gates which belong to the same cluster need to be placed together [1]. Perhaps, each of the clusters is gated by a separate sleep transistor and the sizing of that sleep transistor is generally done by considering the amount of current flowing through the cluster [2].

As per one of the traditional approaches, for the purpose of grouping the logic gates into different clusters, the critical path for the circuit is determined, and according to that, the logic gates which reside in the critical path have been grouped together to form a cluster (C_cluster) and that cluster is generally power gated by a larger sleep transistor. However, the rest of the logic gates can be grouped in one (or, more than one) non-critical cluster (s). The non-critical cluster (NC_cluster) is generally power gated by a regular size sleep transistor [4].

### 3.3 Distributed Sleep Transistor Network

Distributed Sleep Transistor Network is one of the popular means of Power Gating, where the area requirement is found to be much lesser compared to the CBSTD. Conventionally, in case of DSTN, a regular sized sleep transistor has to be placed locally for each of the clusters. And due to the proximity of the sleep transistors, the routing area overhead as well as the wire size become much smaller compared to those for any cluster based design structure [2].

Moreover, considering the 'timing-driven' placement, it is required that the gates with logic connections are placed closed to each other such that the overall interconnect delay gets minimized [2]. Now, the DSTN, as described in previous, can further be advantageous as because of its compatibility with the 'timing-driven' placement.

### 3.4 Cluster Based Tunable Sleep Transistor Cell Power Gating

As reported in [4], the architecture of the tunable sleep transistor cell consists of 4 different sized parallel sleep transistors, which are driven by dedicated control NAND gates. Besides, the outputs of the NAND gates are distributed to the 'Gate' terminals of the sleep transistors through an inverter chain. In this work, we have mainly modified the architecture of the tunable sleep transistor cell of [4], to a simpler structure, governed by (9). Apart from that, here we have used AND gates instead of NAND gates, thereby excluded the use of the separate inverting buffer chain. As shown in Fig. 2, the AND gates receive a 4-bit pattern (B3, B2, B1, B0), and depending upon the SLPBAR1 signal, the values of those 4-bit can be used for the purpose of switching 'ON' or, 'OFF' any of the four sleep transistors. Now, W=135 nm being regular width of the sleep transistors, that we have used in our design, the size of the other three transistors forming the tunable cell can be found from the equation,

$$W_{ST} = \sum_{i=1}^{4} (i*W) \times B_{i-1}, \qquad (9)$$

where, $B_{i-1}$ is meant for obtaining the individual bit values of the 4-bit pattern, and $i$ is for denoting the scaling factor.

As it is illustrated in Fig. 2, the sleep transistor widths are taken as 135 nm, 270 nm, 405 nm, and 540 nm for designing the tunable cell. Besides, the tunable sleep transistor cell has only been used for the C_cluster, whereas the other NC_clusters have been power gated with the non-tunable sleep transistors having smaller sizes (270 nm).

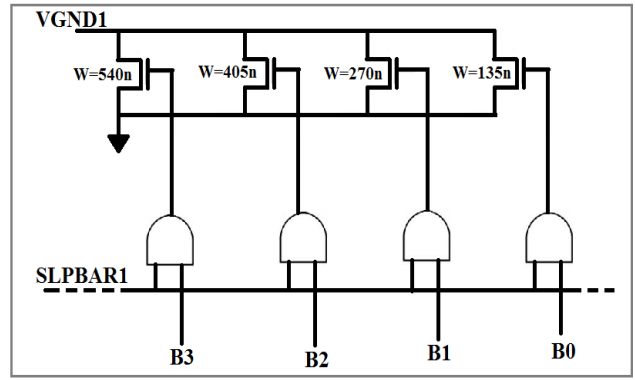

Figure 2. Architecture of the modified tunable sleep transistor cell

## 4. Architecture of the 4×4 multiplier

An extensive analysis has been done here, in this work, with the aim of finding a suitable Power Gating strategy, which can effectively be used in reducing the standby mode leakage power of a digital circuit. For that very purpose, we have actually considered the conventional 4×4 multiplier circuit [8], and applied various Power Gating techniques to gate the circuit. Now, the multipliers, which are vastly used in microprocessors, DSP and communication applications [10, 16], can be simply viewed as the collection of adders [8]. The circuit of the 4×4 multiplier, as shown in Fig. 3, uses a straightforward approach to accumulate the partial

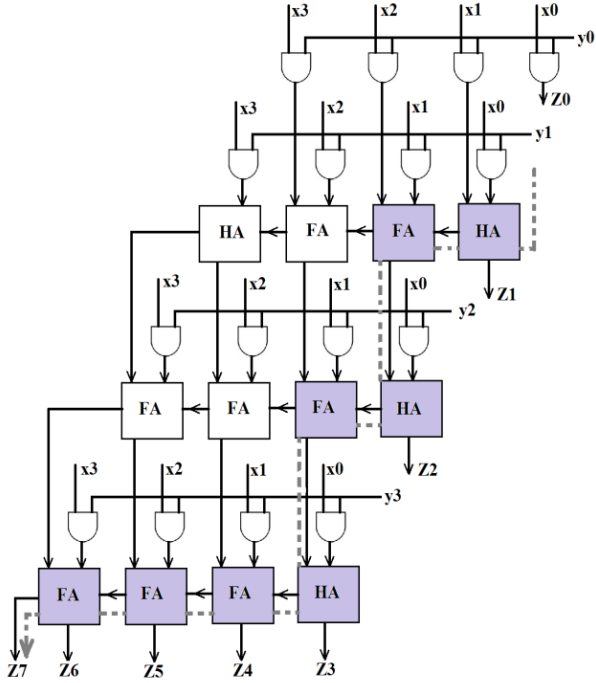

Figure 3. Circuit design of the 4×4 multiplier [8]

products with the help of an array formed by number of adders [8].

Now, for the performance optimization, it is very much required to find out the critical path of the circuit. The dotted line highlighted in Fig. 3, shows the critical path that we have considered in our work [8]. Moreover, while designing the two-bit AND gates, as well as the adder circuits (both full adder and half adder), we have utilized the concept of McCMOS technique (as described in section 2). To optimize the power consumption, as well as to maintain the performance of the circuit, the L and W values of the transistors used for those basic building blocks, are required to be modified.

## 5. Results and Discussions

From Table I, we can have the quantitative information regarding the effects of sleep transistor sizing (in the case of conventional Power Gating scheme) on the performance of the 4×4 multiplier circuit. As it is illustrated in Table I, the gate length has been kept same in all the cases; whereas the width of the sleep transistor has been varied from a nominal value of W=135 nm, to some higher values.

In order to limit the IR drop across the sleep transistor to a certain value (as per the constraint mentioned before, that is the $v_{ST}$ should not exceed 10 % of $V_{dd}$), we have considered the case where W= 700 nm, $v_{ST}$ = 89 mVolt, and the corresponding delay at output = $2.6052 \times 10^{-10}$ second. For the rest of this article, we will refer this delay value as the best case delay ($d_{BC}$).

A similar analysis, for the 4×4 multiplier design Power Gated with CBSTD, has been shown in Table II. However, one more constraint has been included in this case, and according to that, a 10 % increase in delay from the $d_{BC}$ value is taken as the maximum tolerance [4]. As shown in Table II, for a value of W= 400 nm, the maximum delay at output is $2.8091 \times 10^{-10}$ second, which is lesser than the critical value of $2.8657 \times 10^{-10}$ second (i.e., 1.10 times of $d_{BC}$).

Table I. Performance of the 4×4 multiplier circuit, when power gated with the conventional Power Gating scheme

| W (nm) | L (nm) | delay (second) | $v_{ST}$ (mVolt) | $\Delta d / d$ (%) |
|---|---|---|---|---|
| 135 | 45 | $3.4112 \times 10^{-10}$ | 250 | 43.11 |
| 270 |  | $2.9149 \times 10^{-10}$ | 173 | 22.28 |
| 400 |  | $2.7531 \times 10^{-10}$ | 134 | 15.50 |
| 540 |  | $2.6676 \times 10^{-10}$ | 108 | 11.91 |
| 700 |  | $2.6052 \times 10^{-10}$ | 89 | 9.29 |

Table II. Performance of the 4×4 multiplier circuit design, when power gated with the CBSTD technique

| W (nm) | L (nm) | delay (second) | $v_{ST}$ (mVolt) | Shift from the $d_{BC}$ value (%) |
|---|---|---|---|---|
| 100 | 45 | $3.1152 \times 10^{-10}$ | 206 | 19.57 |
| 135 |  | $3.0060 \times 10^{-10}$ | 172 | 15.38 |
| 270 |  | $2.8680 \times 10^{-10}$ | 111 | 10.08 |
| 400 |  | $2.8091 \times 10^{-10}$ | 86 | 7.82 |

Next, for the purpose of comparison of the different Power Gating strategies, the same 4×4 multiplier architecture has been gated with the DSTN Power Gating, as well as with the cluster based tunable sleep transistor cell design. And the results obtained considering the different performance parameters, like power consumption, delay etc., are listed in Table III. All the simulation results are obtained for the operating frequency of 200 MHz and the supply voltage of 1 Volt.

Now in case of DSTN, considering the 'timing driven placement' technique, the entire circuit (of Fig. 3) is divided into seven different rows, and initially, a regular sized sleep transistor (W=135 nm) has been placed locally for each of these rows. Then, we have upsized the width of the sleep transistors and taken into account the $v_{ST}$ values at different tapping points of the DSTN, until those become lesser than the 100 mVolt (which is actually 10 % of $V_{dd.}$). It is observed that for the W = 270 nm, all the tapping points (except only one) meet the aforesaid constraint; whereas the maximum delay at output of the multiplier circuit, is also found to be lesser than the critical value of $2.8657\times10^{-10}$ second.

For the case of cluster based tunable sleep transistor cell Power Gating, the tunable cell architecture consists of 4 different sized parallel sleep transistors. And those sleep transistors are actually driven by the dedicated control AND gates which receive a 4-bit pattern B3, B2, B1, B0. For comparing the performance of this Power Gating technique with DSTN Power Gating (as shown in Table III ), the bit configuration word is set to "1000" which is the nominal case.

Fig. 4 shows the variation in virtual rail voltage (VGND1 of Fig. 2) with the change in bit configuration. Whereas, for all the possible bit configurations, the different values of the Average Power consumption of the 4×4 multiplier circuit are illustrated in Fig. 5. Average Power consumption increases from $1.34026\times10^{-5}$ Watt to $1.37476\times10^{-5}$ Watt, as the bit pattern varies from "0001" to "1111"; however, at the same time, the value of maximum delay at output decreases from $2.8174\times10^{-10}$ second to $2.5054\times10^{-10}$ second.

Now, compared to the 4×4 multiplier with DSTN (as shown in Table III), though the same with cluster based tunable sleep transistor cell Power Gating consumes almost similar power, but looking at the other aspects it provides much better performance. Again, for the sake of comparison, if we consider the 4×4 multiplier circuit of Fig. 3, without any Power Gating scheme, then the value of the Average Power and the delay will come as $1.3862\times10^{-5}$ Watt and $2.3836\times10^{-10}$ second. Therefore, this modified tunable sleep transistor cell can obtain a 1.61 % reduction in the Average Power consumption at the cost of 6.79 % increase in delay. But, obviously looking at the performances of the other Power Gating schemes (like, conventional Power Gating, CBSTD, DSTN), the delay provided by the multiplier circuit with tunable sleep transistor cell Power Gating is found to be much lesser.

Table III. Comparison of the modified tunable sleep transistor cell Power Gating technique with the DSTN Power Gating

| Power Gating Strategy | $V_{dd}$ (Volt) | Avg. Power (Watt) | delay (second) | Max. $v_{ST}$ value/ Virtual rail voltage as VGND1 (mVolt) | Improvement in performance, w.r.t. the $d_{BC}$ value (%) |
|---|---|---|---|---|---|
| DSTN (W = 270 nm) | 1.0 | $1.3556\times10^{-5}$ | $2.5871\times10^{-10}$ | 114 | 0.9 |
| modified tunable sleep transistor cell (for the bit pattern of "1000") | 1.0 | $1.3638\times10^{-5}$ | $2.5455\times10^{-10}$ | 74 | 2.29 |

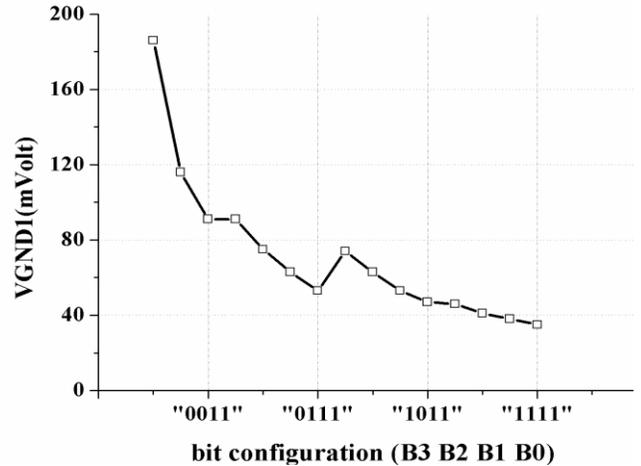

Figure 4. Virtual rail voltage versus bit configuration

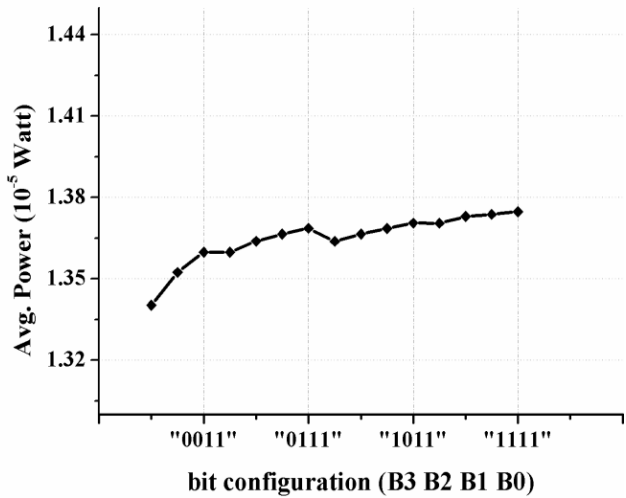

Figure 5. Average Power consumption details for all the possible bit configurations

## 6. Conclusion

In this work, we have focused on the impact of the several Power Gating strategies which significantly reduces the standby mode leakage power in any CMOS circuit, while maintaining a desirable performance or, speed. A fair comparison looking at the performances of the 4×4 multiplier circuit with the introduction of the different Power Gating schemes such as conventional Power Gating, CBSTD, DSTN, and cluster based tunable sleep transistor cell Power Gating, has been presented here. Compared to DSTN, as well as the other Power Gating schemes as discussed, the cluster based tunable sleep transistor cell Power Gating can provide best case performance with a 2.29% improvement with respect to the $d_{BC}$. Moreover, the tunable sleep transistor cell has its inherent advantage of having the programmable parallel connection of transistors, which leads to the maximum dynamicity.

ACKNOWLEDGMENT

Authors would like to thank SMDP-II project lab., IC Design & Fabrication Centre, Jadavpur University for getting the opportunity to carry out this work using SPICE Tools .